\documentclass[aps,prc,preprint,groupedaddress,showpacs,showkeys,nofootinbib]{revtex4}
\usepackage{epsfig,psfig,float,amssymb,latexsym}
\newcommand{\be}{\begin{equation}}
\newcommand{\ee}{\end{equation}}
\newcommand{\ba}{\begin{eqnarray}}
\newcommand{\ea}{\end{eqnarray}}

\begin{document}

\preprint{IFIC$-$02$-$0529}
\preprint{FTUV$-$02$-$0529}

\title{Phi meson mass and decay width in nuclear matter}

\author{D. Cabrera}
\email[]{Daniel.Cabrera@ific.uv.es}
\affiliation{Departamento de F\'{\i}sica Te\'orica and IFIC, 
Centro Mixto Universidad de Valencia-CSIC, 
Institutos de Investigaci\'on de Paterna, Apdo. correos 22085, 
46071, Valencia, Spain}
\author{M.J. Vicente Vacas}
\email[]{Manuel.J.Vicente@ific.uv.es}
\affiliation{Departamento de F\'{\i}sica Te\'orica and IFIC, 
Centro Mixto Universidad de Valencia-CSIC, 
Institutos de Investigaci\'on de Paterna, Apdo. correos 22085, 
46071, Valencia, Spain}

\date{\today}

\begin{abstract}
The $\phi$ meson spectrum, which in vacuum is dominated by its coupling
to the $\bar{K} K$ system, is modified in nuclear matter. Following a model
based on chiral $SU(3)$ dynamics we calculate the $\phi$ meson selfenergy in
nuclear matter considering the $K$ and $\bar{K}$ in-medium properties. For the
latter we use the results of previous calculations which account for $S-$ and
$P-$wave kaon-nucleon interactions based on the lowest order meson-baryon
chiral effective Lagrangian, and this leads to a dressing of the kaon
propagators in the medium. In addition, a set of vertex corrections is
evaluated to fulfill gauge invariance, which involves contact couplings of the
$\phi$ meson to $S-$wave and $P-$wave kaon-baryon vertices. Within this scheme
the mass shift and decay width of the $\phi$ meson in nuclear matter are
studied.
\end{abstract}

\pacs{13.25.-k, 14.40.Cs, 25.80.-e}
\keywords{$\phi$ decay in nuclei, Chiral Lagrangian, $\bar{K}$-nucleus
interaction}

\maketitle

\section{\label{s1}Introduction}
The study of the renormalization of the hadron properties (like effective
masses and widths) in a hot/dense baryonic medium has been a matter of wide
interest during the last years (see for instance ref. \cite{Rapp:2000ej}).
Electromagnetic decays of vector mesons are specially suited to explore the
high densities which occur in nuclei or neutron stars, since the emitted
dilepton pairs, which carry the spectral information of the vector meson in the
moment of the decay, undergo a low distortion before they abandon the dense
medium and reach the detector. In particular, the $\phi$ meson stands as a a
unique probe for a possible partial restoration of chiral symmetry in hadronic
matter, since it does not overlap with other light resonances in the mass
spectrum. In addition, the modification of the $\phi$ properties in the medium
is strongly related to the renormalization of the kaon properties in the
medium, a topic which has also received much attention because of notable
deviations from the low density theorem needed to reproduce kaonic atoms data
\cite{Batty:1997zp,Hirenzaki:2000da,Baca:2000ic,Gal:2001da} and the possibility
of formation of kaon condensates in neutron stars \cite{Kaplan:1986yq}. The
$\phi$ properties in a nuclear medium could be tested through detection of its
decay products in $A-A$ and $p-A$ collisions \cite{Pal:2002aw}, by means of the
reaction $\pi^- p \to \phi n$ in nuclei \cite{Klingl:1998tm} or the recently
proposed method based on inclusive $\phi$ photoproduction in nuclei
\cite{Oset:2001na}.

Concerning the change of the $\phi$ properties in the medium, much of the work
done has been partly stimulated by the idea that partial restoration of chiral
symmetry may modify the masses of mesons at finite temperature and/or density.
The $\phi$ mass change has been studied in several approaches like using
effective Lagrangians 
\cite{Klingl:1998tm,Kuwabara:1995ms,Song:1996gw,Bhattacharyya:1997kx,Klingl:1997kf},
QCD sum rules \cite{Asakawa:1994tp,Zschocke:2002mn} or the Nambu-Jona-Lasinio
model  \cite{Blaizot:1991af}. The $\phi$ width modification in matter has also
been subject of study in a dropping meson mass scenario
\cite{Bhattacharyya:1997kx,Ko:tp,Shuryak:1992yy,Lissauer:1991fr,Panda:1993ik,Ko:1994id},
as a result of collisional broadening through $\phi$-baryon \cite{Smith:1998xu}
or $\phi$-meson \cite{Alvarez-Ruso:2002ib} scattering  processes and on the
basis of modifications of the main decay channels in vacuum
\cite{Klingl:1998tm,Oset:2001eg}. All these works point at a sizeable
renormalization of the $\phi$ width and a small mass shift\footnote{Except for
ref. \cite{Asakawa:1994tp} in which a mass shift of a few hundred MeV is
reported under hot and dense matter conditions.}.

The purpose of this work is to obtain the $\phi$ meson selfenergy in cold
symmetric nuclear matter from which we shall study how the mass and decay width
are modified in the medium. We follow the lines of ref. \cite{Oset:2001eg} in
which the main input to the calculation of the $\phi$ selfenergy in nuclear
matter is the kaon selfenergy of \cite{Ramos:2000ku}, arising from $S-$ and
$P-$wave interactions with the nucleons.  The $S-$wave kaon selfenergy is
obtained from a selfconsistent coupled channel unitary calculation of the
$\bar{K}N$ amplitude based on effective chiral Lagrangians, in which Pauli
blocking,  the renormalization of the pions and mean field potentials of the
baryons in the intermediate states are taken into account. The $P-$wave kaon
selfenergy is driven by the coupling to hyperon-hole excitations. Although we
adhere to this approach for the kaon selfenergy, we introduce some technical
novelties. Whereas the work in \cite{Oset:2001eg} is devoted to the calculation
of the in-medium $\phi$ width and thus only the imaginary part of the $\phi$
selfenergy is obtained, we also focus on the real part to look for the shift in
the $\phi$ mass. This leads us to improve the model of \cite{Oset:2001eg} by
including additional mechanisms required by gauge invariance as well as
performing a finer study of the relativistic recoil corrections considered in
the $P-$wave kaon selfenergy. In section \ref{s2} we detail our model of the $\phi$
selfenergy for which we adhere to the vector field representation of ref.
\cite{Klingl:1998tm}. $S-$ and $P-$wave kaon selfenergies are quoted as the
major input of the calculation in nuclear matter and a set of vertex
corrections involving contact $\phi$-meson-baryon vertices are considered as
requested by gauge invariance. In section \ref{s3} we present our results for the
$\phi$ selfenergy and the consequent findings for the $\phi$ mass shift and
width, and summarize our conclusions.

\section{\label{s2}$\phi$ meson selfenergy in vacuum and in nuclear matter}

In this work we shall deal with the coupling of the $\phi$ to $K \bar{K}$
channels, which in vacuum accounts for 85 \% of the total decay width. It was
found in \cite{Klingl:1998tm,Oset:2001eg} that the in medium $\bar{K} K$
related channels give rise to a $\phi$ width several times larger than in free
space. Here we shall focus in the medium modifications of the whole $\phi$
meson selfenergy (both real and imaginary parts) arising from $\bar{K} K$
related channels, and we shall ignore further renormalization coming from other
contributing channels in free space as the $3\pi$ channel.

\begin{figure}
\includegraphics[width=0.5\textwidth]{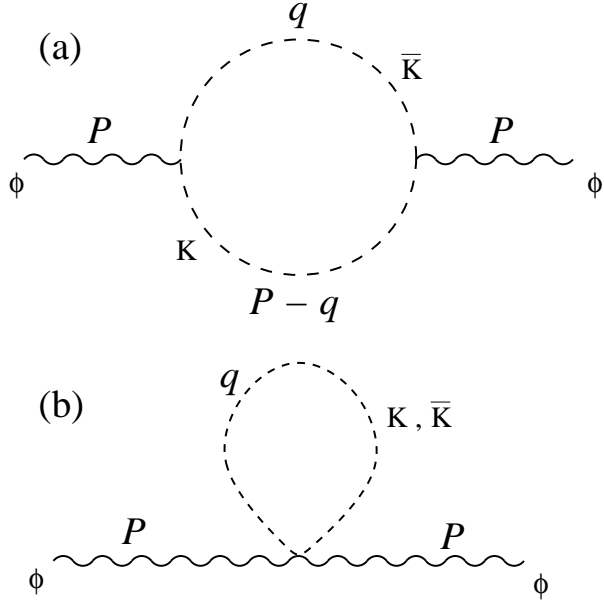}
\caption{\label{vac_self}Diagrams contributing to the ${\cal O} (g_{\phi}^2)$
$\phi$ selfenergy in free space: (a) two-kaon loop, (b) kaon tadpole.}
\end{figure}

To describe the interactions of the $\phi$ meson with pseudoscalar mesons and
baryons, we follow the model developed in \cite{Klingl:1997kf,Klingl:1996by}.
The Lagrangian describing the coupling of the $\phi$ meson to kaons reads 
\ba \label{Lagr1}
{\cal L}_{\phi,kaons} = -i g_{\phi} \phi_{\mu} (K^-
\partial^{\mu} K^+ - K^+ \partial^{\mu} K^- + \bar{K}^0
\partial^{\mu} K^0 - K^0 \partial^{\mu} \bar{K}^0)
\nonumber \\
+ g_{\phi}^2 \phi_{\mu} \phi^{\mu} (K^- K^+ + \bar{K}^0 K^0) ,
\ea
and it provides, at ${\cal O} (g_{\phi}^2)$, a $\phi$ meson selfenergy
$\Pi_{\phi}^{free}$ consisting of two-kaon loop contributions and kaon tadpole
terms (see Fig. \ref{vac_self}). The free width to the $K \bar{K}$ channel can
be readily calculated from the imaginary part of $\Pi_{\phi}^{free}$ and, by
comparing to the experimental value of $\Gamma_{\phi \to K^+ K^-}$ one gets
$g_{\phi} = 4.57$.

\begin{figure}
\includegraphics[width=0.5\textwidth]{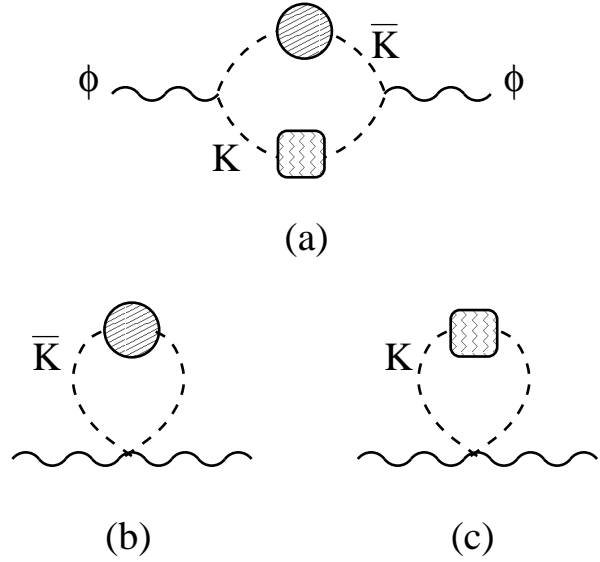}
\caption{\label{med_self}Medium modified diagrams contributing to the $\phi$ meson
selfenergy in nuclear matter after including kaon selfenergies. The dressing of
$\bar{K}$($K$) propagator is indicated by a round (squared) box.}
\end{figure}

The calculation of the $\phi$ meson selfenergy in nuclear matter,
$\Pi_{\phi}^{med}$, proceeds by a dressing of the kaon propagators including
both $S-$ and $P-$wave selfenergies, as well as vertex corrections demanded by
gauge invariance. We perform a subtraction of the free selfenergy
$\Pi_{\phi}^{free}$ from the in-medium selfenergy $\Pi_{\phi}^{med}$ which, as
we shall discuss below, provides a convergent result. From this in-medium
subtracted selfenergy, $\Delta \Pi_{\phi}^{med} = \Pi_{\phi}^{med} -
\Pi_{\phi}^{free}$, the width and mass shift of the $\phi$ meson will be
obtained in section \ref{s3}. Since we are interested in both the real and imaginary
parts of the $\phi$ selfenergy, the tadpole terms have to be explicitly
included in the calculation and properly modified in the medium.  The first
step to include the effects of the medium consists of a modification of the
kaon propagator with a proper selfenergy $\Pi_K (q^0,\vec{q};\rho)$. A
diagrammatic representation is shown in Fig. \ref{med_self}.
After this is done, the kaon propagator is given, in its spectral
representation \cite{Klingl:1998tm}, by
\be
\label{dressed_prop}
D_{\bar{K} (K)} (q^0,\vec{q};\rho) =
\int_0^{\infty} d\omega \bigg ( \frac{S_{\bar{K} (K)}
(\omega,\vec{q};\rho)}{q^0-\omega+i\eta} - \frac{S_{K (\bar{K})}
(\omega,\vec{q};\rho)}{q^0+\omega-i\eta} \bigg ) ,
\ee
where $S_{\bar{K} (K)}$ is the spectral function of the $\bar{K} (K)$ meson,
\be
\label{spectral_kaons}
S_{\bar{K} (K)}(q^0,\vec{q};\rho) = - \frac{1}{\pi}
\frac{\textrm{Im} \, \Pi_{\bar{K} (K)}(q^0,\vec{q};\rho)}
{| (q^0)^2 - \vec{q}\,^2 - m_K^2 - \Pi_{\bar{K} (K)}(q^0,\vec{q};\rho)|^2}
\ee
and $\Pi_{\bar{K} (K)}$ is the $\bar{K} (K)$ selfenergy in nuclear matter. 
From now on we take the isospin limit and
set $m_K=m_{\bar{K}}=(m_{K^+}+m_{K^0})/2$.
The in-medium $\phi$ selfenergy arising from dressing the
kaon propagators for a $\phi$ meson at rest reads
\ba
\label{self_only_dressing}
\Pi_{\phi}^{ij} (P^0;\rho) = \delta^{ij} i 2
g_{\phi}^2 \frac{4}{3} \int  \frac{d^4q}{(2\pi)^4} \vec{q}\,^2 D_K(P-q;\rho)
D_{\bar{K}}(q;\rho) \nonumber \\ + \delta^{ij} i 2 g_{\phi}^2 \bigg \lbrace
\int \frac{d^4q}{(2\pi)^4} D_{\bar{K}}(q;\rho) + \int \frac{d^4q}{(2\pi)^4}
D_K(q;\rho) \bigg \rbrace \equiv \delta^{ij} \Pi_{\phi} (P^0;\rho),
\ea
where only spatial components are non-vanishing.
If we take a constant selfenergy for the kaon, as in
\cite{Ramos:2000ku,Kaiser:1995eg}, the $dq^0$ and $d\Omega_q$ integrations can
be readily evaluated and the $\phi$ selfenergy takes the simple form
\ba
\label{self_only_dressing_Lehm}
\Pi_{\phi} (P^0;\rho) = 2 g_{\phi}^2 \frac{1}{2\pi^2}\frac{4}{3} \int_0^{\infty}
dq \, \vec{q}\,^2 \bigg \lbrace \frac{\vec{q}\,^2}{\widetilde{\omega}(q)}
\int_0^{\infty} d\omega \frac{S_{\bar{K}}(\omega,|\vec{q}|;\rho) \, (\omega +
\widetilde{\omega}(q))}
{(P^0)^2 - (\omega + \widetilde{\omega}(q))^2 + i\epsilon}
\nonumber \\
+ \frac{3}{4} \bigg \lbrack \int_0^{\infty} d\omega
S_{\bar{K}}(\omega,|\vec{q}|;\rho) + \frac{1}{2 \widetilde{\omega}(q)} \bigg \rbrack
\bigg \rbrace ,
\ea
where $\widetilde{\omega}(q)^2 = \vec{q}\,^2 + m_K^2 + \Pi_K(\rho)$.

In the following sections we discuss how the kaon selfenergies are obtained. We
follow closely ref. \cite{Oset:2001eg} and focus only on the differences
introduced in this work.

\subsection{$S-$wave kaon selfenergy}
The interactions of $K^+$, $K^0$ and $\bar{K}^0$, $K^-$ with the nucleons of
the medium are rather different and it is necessary to treat them separately.
Since there are no $S=1$ baryonic resonances the $KN$ interaction is smooth at
low energies. We take the approach of \cite{Ramos:2000ku,Kaiser:1995eg} where a
$t \, \rho$ approximation is used and a constant selfenergy is obtained.

The $\bar{K} N$ interaction, however, is dominated at low energies by the
excitation of the $\Lambda (1405)$ resonance, which appears just below the
$\bar{K} N$ threshold. To account for a realistic $\bar{K}$ selfenergy in
nuclear matter we follow the coupled channel chiral unitary approach to $S-$wave
$\bar{K} N$ scattering developed in \cite{Oset:1998it}. Starting from this
successful scheme in free space, an effective interaction in the medium was
obtained in \cite{Ramos:2000ku} solving a coupled channel Bethe-Salpeter
equation including Pauli blocking on the nucleons, mean-field binding
potentials for the baryons involved and the medium modifications of $\pi$
mesons and $\bar{K}$ mesons themselves, leading thus to a selfconsistent
calculation. By summing over the occupied states in the Fermi sea, the
$\bar{K}$ spectral function was obtained. The inclusion of the $S-$wave
selfenergy accounts for decay channels of the $\phi$ meson of the type $K Y M
h$, the major contribution coming from $Y=\Sigma$, $M=\pi$.  The range of
validity of this model covers kaon energies and momenta up to 1.5 GeV. Outside
this range we take a constant value matching it to the boundary value at energy
1.5 GeV and zero momentum. We have checked that the dependence of our results
on the choice of the matching point is negligible. We refer the reader to refs.
\cite{Ramos:2000ku,Oset:1998it} for details of the model.

\subsection{$P-$wave kaon selfenergy}
Kaons from $\phi$ decay are emitted with rather low energies, what makes the
$S-$wave selfenergy a major renormalization source to be included in nuclear
matter. However, in \cite{Klingl:1998tm,Oset:2001eg} it was noticed that
including $P-$wave kaon couplings to nucleons gives rise to a considerable
source of renormalization of the $\phi$ in the medium, leading to a further
increase of the $\phi$ width of the same order as that provided by the $S-$wave
contribution.

\begin{figure}
\includegraphics[width=0.7\textwidth]{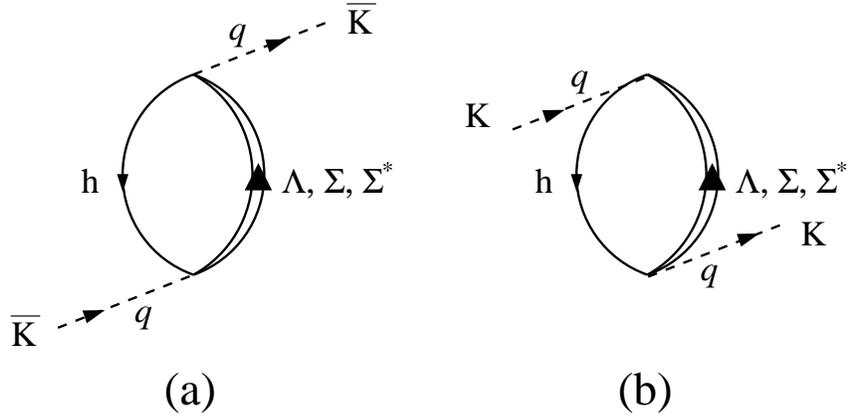}
\caption{\label{kaon_pwave}Kaon $P-$wave selfenergy diagrams: (a) $\bar{K}$ direct
term; (b) $K$ crossed term.}
\end{figure}

We include $P-$wave kaon selfenergies accounting for the excitation of $\Lambda
h$, $\Sigma h$ and $\Sigma^* h$ pairs ($\Sigma^*=\Sigma^*(1385)$). The
corresponding many-body mechanisms are shown in Fig. \ref{kaon_pwave}. Notice
that, because of strangeness conservation, only direct terms are permitted for
the $\bar{K}$ excitations. Conversely, the $K$ selfenergy arises from the
crossed terms. Since the excited hyperon is far off-shell in the crossed
kinematics, we expect the $K$ to be barely modified by the $P-$wave selfenergy
and we shall not include it in the calculation.
The $\bar{K}NY$ vertices involved are derived from the lowest order chiral
Lagrangian coupling the octet of pseudoscalar mesons to the $1/2^+$ baryon
octet \cite{Ramos:2000ku} and then treated in a non-relativistic approach. The
$P-$wave $\bar{K}$ selfenergy then reads
\ba
\label{kpw}
\Pi^{Pwave}_{\bar{K}}(q^0,\vec{q};\rho) &\equiv&
\vec{q}\,^2 \widetilde{\Pi}^{Pwave}_{\bar{K}}(q^0,\vec{q};\rho)
\nonumber \\
&=& \frac{1}{2} \widetilde{V}^2_{K^- p \Lambda}
f_\Lambda^2(q) \vec{q}\,^2 U_\Lambda(q^0,\vec{q};\rho)
\nonumber \\
&+& \frac{3}{2} \widetilde{V}^2_{K^- p \Sigma^0}
 f_\Sigma^2(q) \vec{q}\,^2 U_\Sigma(q^0,\vec{q};\rho)
\\
&+& \frac{1}{2} \widetilde{V}^2_{K^- p \Sigma^{*0}}
 f_{\Sigma^*}^2(q) \vec{q}\,^2 U_{\Sigma^*}(q^0,\vec{q};\rho) \ ,
\nonumber
\ea
where $U_Y$ ($Y=\Lambda,\Sigma,\Sigma^*$) stands for the Lindhard function
including only direct kinematics, and the $\widetilde{V}_{\bar{K}NY}$ factors
contain the required $\bar{K}NY$ couplings and isospin factors which can be 
found in ref. \cite{Oset:2001eg}. 

In the model of ref. \cite{Oset:2001eg} a
non-relativistic treatment of the $\bar{K}NY$ vertices is used, including
relativistic recoil corrections of ${\cal O} (1/M_Y)$, to deal with the
imaginary part of the $\phi$ selfenergy. However, in this work, we are also
concerned with the calculation of $\textrm{Re} \, \Pi_{\phi}^{med}$ and larger
energies and momenta are involved in the loop contributions. We have improved
the model of \cite{Oset:2001eg} in the treatment of the relativistic recoil
corrections, quoted as the $f_Y^2(q)$ functions in eq. (\ref{kpw}). Starting
from the fully relativistic $\bar{K}N \to Y \to \bar{K}N$ amplitude with an
approximately at rest in-medium nucleon and keeping the positive energy
intermediate hyperon propagator, we get the following correction factor,
\be
\label{RF_B}
f_Y^2(q) = 1 + \frac{(E_Y-M_Y)^2 - (q^0)^2}{4 E_Y M_Y} ,
\ee
where $E_Y=\sqrt{M_Y^2 + \vec{q}\,^2}$ and $q^0$ is the kaon energy.

Additionally, to account for
the off-shell behaviour of the kaon-baryon vertices we have included static
dipolar form factors.  In \cite{Klingl:1998tm,Oset:2001eg} relativistic dipolar
form factors where used, each $\bar{K}NY$ vertex carrying a $(\Lambda^2 /
(\Lambda^2-(q^0)^2+\vec{q}\,^2))^2$ factor with $\Lambda=1.05$ GeV. In this
work, the use of such form factors is not a proper approach because it
generates fictitious poles in the loop contributions to the $\phi$ selfenergy.
Instead, we choose to use a static version.

 To compensate for both changes, we properly rescale the $\bar{K}NY$  form
factors to keep consistency with the vertices used in
\cite{Klingl:1998tm,Oset:2001eg}. Therefore we multiply each $\bar{K}NY$ vertex
by $C \, (\Lambda / (\Lambda + \vec{q}\,^2))^2$ and we fix the $C$ constant by
performing an on-shell matching to the expressions in
\cite{Klingl:1998tm,Oset:2001eg}. Details on how this matching is done are
given in the appendix \ref{Ap1}.

\subsection{Vertex corrections and gauge invariance}
The introduction of the $\phi$ coupling to kaons and $\bar{K}NY$ interactions
in a gauge vector description of the $\phi$ meson generates additional $\phi
\bar{K}NY$ contact interaction vertices.
These vertices
can be evaluated systematically as done in ref. \cite{Klingl:1997kf}.
Alternatively, one can also obtain the Feynman rules for these vertices by
requiring the $\phi N \to KY$ amplitude to vanish when replacing the vector
meson polarization by its four-momentum.

\begin{figure}
\includegraphics[width=0.5\textwidth]{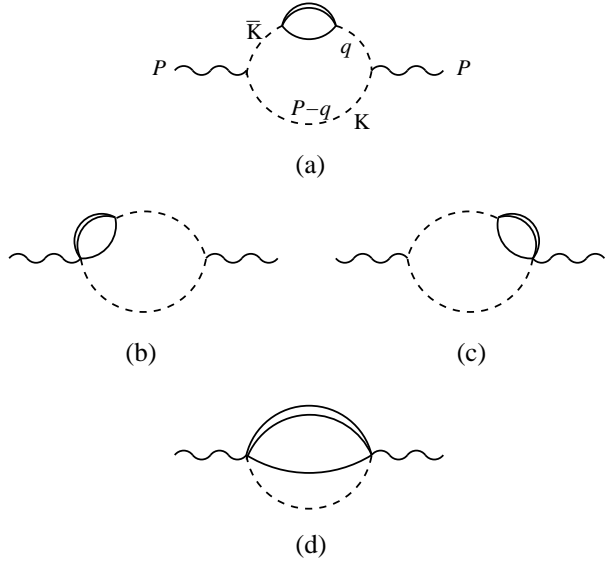}
\caption{\label{vc_graphs}${\cal O}(\rho)$ contributions to the $\phi$ selfenergy
from $P-$wave kaon selfenergy insertions and related vertex
corrections.}
\end{figure}

At the lowest order in the nuclear density, $P-$wave kaon selfenergy insertions
plus vertex corrections are considered in the diagrams depicted in Fig.
\ref{vc_graphs}.
The contribution of diagram (a) in Fig. \ref{vc_graphs} has already been
considered in eqs. (\ref{self_only_dressing}), (\ref{self_only_dressing_Lehm})
since it corresponds to a single bubble insertion in the $\bar{K}$ propagator
and it is accounted for by using fully dressed kaon propagators in the
calculation of the $\phi$ meson selfenergy. Diagrams (b-d), however, cannot be
cast as medium modifications of the meson propagators but as genuine vertex
corrections. Let us call $\Pi_{\phi}^{VC1}$ and $\Pi_{\phi}^{VC2}$ to the
selfenergy contributions associated to diagrams (b)+(c) and (d), respectively.
Using the Feynman rules derived in \cite{Oset:2001eg} one finds, at first order
in density,
\be
\label{vc1}
\Pi_{\phi}^{VC1} (P^0;\rho) = i 2 g_{\phi}^2 \frac{4}{3} \int
\frac{d^4q}{(2\pi)^4} \vec{q}\,^2 \widetilde{\Pi}^{Pwave}_{\bar{K}} (q;\rho)
\frac{1}{(P^0-q^0)^2-\omega^2(q)+i\epsilon}
\frac{1}{(q^0)^2-\omega^2(q)+i\epsilon},
\ee
\be
\label{vc2}
\Pi_{\phi}^{VC2} (P^0;\rho) = i 2 g_{\phi}^2 \int \frac{d^4q}{(2\pi)^4}
\widetilde{\Pi}^{Pwave}_{\bar{K}} (q;\rho)
\frac{1}{(P^0-q^0)^2-\omega^2(q)+i\epsilon} ,
\ee
with $\omega^2(q)=\vec{q}\,^2 + m_K^2$. In eqs. (\ref{vc1}), (\ref{vc2}) we
have used bare kaon propagators. These results can be  further extended to
higher orders in density by considering additional selfenergy insertions in the
kaon propagators. Eventually we substitute the bare kaon propagators in
eqs. (\ref{vc1}), (\ref{vc2}) by the fully dressed propagators described in the
previous sections.
Both contributions $\Pi_{\phi}^{VC1}$, $\Pi_{\phi}^{VC2}$
are finite provided that suitable form factors, as discussed in the previous
section, are used in the $P-$wave $\bar{K}NY$vertices.

\begin{figure}
\includegraphics[width=0.5\textwidth]{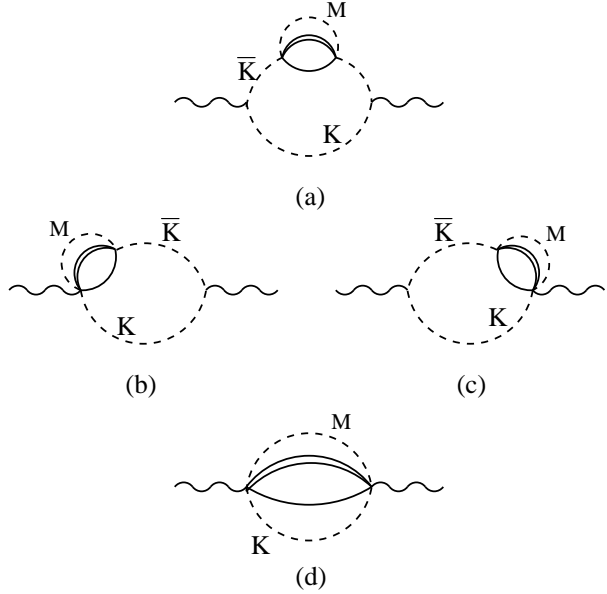}
\caption{\label{vc_graphsII}${\cal O}(\rho)$ contributions to the $\phi$ selfenergy
from $S-$wave kaon selfenergy insertions and related vertex
corrections.}
\end{figure}

In the same line as done for the kaon-nucleon $P-$wave interaction, gauge
invariance prescribes a coupling of the $\phi$ meson to the $S-$wave
$\bar{K}NMY$ vertices. This coupling is obtained by substituting the ordinary
derivative in the meson-baryon chiral Lagrangian of ref. \cite{Oset:1998it}
(see eq. (5) therein) by a covariant derivative which includes the vector meson
field $SU(3)$ matrix $V^{\mu}$ and reads
\be
\label{covderivat}
\partial^{\mu} \Phi \to D^{\mu} \Phi = \partial^{\mu} \Phi - i \frac{g}{2}
\lbrack V^{\mu} , \Phi \rbrack ,
\ee
where $\Phi$ represents the pseudoscalar meson field matrix (we follow the
notation of \cite{Klingl:1997kf} and therefore $g_{\phi} = g / \sqrt{2}$). It
is also possible to obtain the actual Feynman rules calculating the amplitude
$t_{MY \to \bar{K}N\phi} \equiv t_{MY \to \bar{K}N\phi}^{\nu}
\epsilon_{\nu}(\phi)$, and requiring $t_{MY \to \bar{K}N\phi} ^{\nu} P_{\nu} =
0$ with $P$ the four-momentum carried by the $\phi$ meson field. The $\phi$
selfenergy diagrams involving $S-$wave kaon selfenergy insertions plus related
vertex corrections are shown in Fig. \ref{vc_graphsII}.
As in the case of the $P-$wave gauging, the contribution of the first diagram
in Fig. \ref{vc_graphsII} is already included in eqs.
(\ref{self_only_dressing}), (\ref{self_only_dressing_Lehm}) by using the
renormalized kaon propagators dressed with the $S-$wave selfenergy of ref.
\cite{Ramos:2000ku}. We expect the other three terms (Fig.
\ref{vc_graphsII}(b-d) to give a small contribution to the imaginary part of
the $\phi$ selfenergy because of the reduced phase space of the intermediate
states. We evaluate them in appendix \ref{Ap2} and find both the real and imaginary
contributions to be negligible as compared to the leading contributions to
$\Pi_{\phi}^{med}$.

In addition to the vertex corrections presented in this section one should
include also other contributions in which the $\phi$ meson directly couples to
the hyperons. These terms were also considered in \cite{Cabrera2002} for the
case of the $\rho$ meson selfenergy in the medium. We have checked numerically
that all of them are about two orders of magnitude smaller than the relevant
pieces in the $\phi$ selfenergy, essentially driven by meson selfenergy
insertions and $P-$wave vertex corrections. One of the reasons is that whenever
the $\phi$ is involved in a $\phi YY$ coupling, there is a hyperon propagator
which lies far off-shell in any of these terms.

The introduction of the contact vertices, following a minimal coupling scheme
as in refs. \cite{Klingl:1997kf,Oset:2001eg}, guarantees gauge invariance and
therefore the $\phi$ selfenergy is transverse. In our approach, there are small
violations of gauge invariance due to the omitted diagrams discussed above, the
non-fully relativistic treatment of the vertices involving baryons, which keep
only the lowest orders in a $(p/M)$ expansion \cite{Ramos:2000ku}; and also to
the fact that only the positive energy part of the hyperon propagators has been
included in the calculation. We have explicitly checked that transversality of
the $\phi$ selfenergy is achieved at the lowest order in a ($p/M$) expansion,
where $p$ is the kaon momentum, when all the diagrams in Figs. 4 and 5, plus
those with the $\phi$ meson coupled to hyperon lines, are considered. In
particular, for the $\phi$ at rest, which is the case under study, this means
that $\Pi_{\phi}^{00}$, $\Pi_{\phi}^{i0}$ and $\Pi_{\phi}^{0j}$ vanish at the
lowest order that contributes to $\Pi_{\phi}^{ij}$ in a $(p/M)$ expansion.
Numerically, we have found that these components are indeed negligible as
compared to $\Pi_{\phi}^{ij}$.

We have also checked that, for a $\phi$ meson at rest, the relevant momenta
running in the kaon loops are around 300 MeV. Thus the corrections to the
non-relativistic approximation, which behave as powers of the kaon momentum
over the hyperon mass, are expected to be small. However, for high $\phi$
momenta, that are easily obtained in a typical heavy ion collision, a fully
relativistic calculation of the interaction vertices and baryon propagators
would be advisable.

\subsection{Regularization}
The $\phi$ meson selfenergy, as introduced in the previous sections, is a
quadratically divergent object either in the medium or in free space, and
therefore a regularization procedure is required to remove divergences.
Actually, since we are interested in the medium modifications of the $\phi$
meson, we calculate a subtracted in-medium selfenergy, $\Delta
\Pi_{\phi}^{med}$, introduced at the beginning of section \ref{s2}. This procedure 
cancels all divergences leading to a finite result both for the real and 
imaginary parts of $\Delta \Pi_{\phi}^{med}$.
To perform the subtraction we calculate $\Pi_{\phi}^{free}$ and
$\Pi_{\phi}^{med}$ in a momentum cut-off scheme, so that each of them
depends on a regularization parameter $q_{max}$, but the subtraction gives 
a finite result in the limit $q_{max} \to \infty$.

Once the vacuum $\phi$ selfenergy is subtracted one is left, at first order in
density, with single meson selfenergy insertion diagrams in the propagators and
vertex correction involving one baryonic loop. Those diagrams involving
$P-$wave couplings are convergent once the form factors are considered, as can
be seen from power counting.  The $S-$wave selfenergy insertion diagrams are
also convergent provided that the high energy-momentum behaviour of the
$S-$wave kaon selfenergy is at most a constant, of course density dependent, as
in our model.

The convergence of the calculation is easily shown by obtaining explicitly the
subtraction of the in-medium $\phi$ selfenergy and the free
$\phi$ selfenergy, and using for the $S-$wave kaon selfenergy its asymptotic
value. In a cut-off regularization scheme, the subtraction reads
\be
\label{med_contribution2}
\Pi_{\phi , \,\, asympt}^{med} - \Pi_{\phi}^{free} =
\mathcal{C}_0 + \mathcal{C}_{-2} \frac{1}{q_{max}^2}
+ {\cal O} (\frac{1}{q_{max}^3}).
\ee
The quadratic and logarithmic pieces cancel exactly and in the limit $q_{max}
\to \infty$ we are left with a finite part $\mathcal{C}_0$ which depends on
$P^0$ and the density.

\section{\label{s3}Results and discussion}
\begin{figure}
\includegraphics[width=0.7\textwidth]{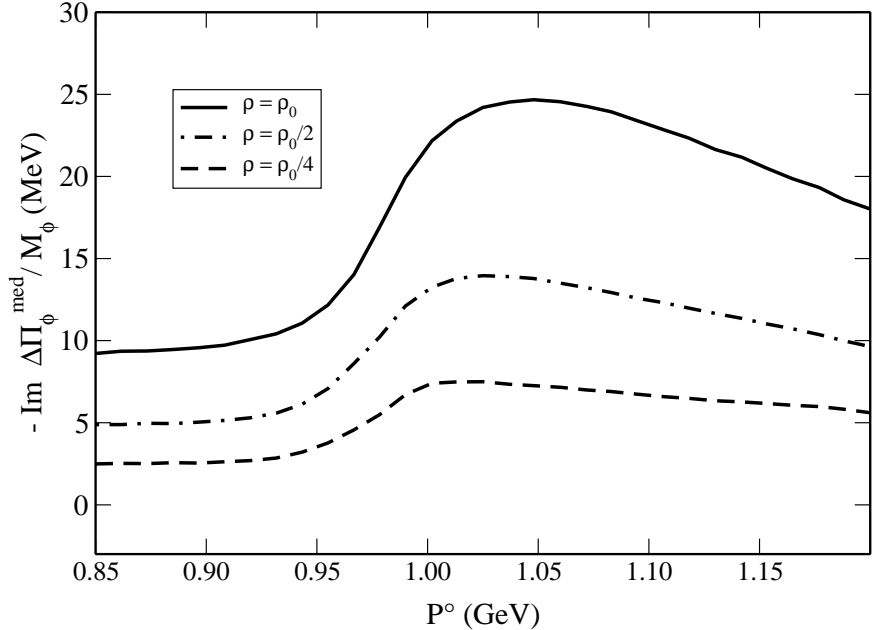}
\caption{\label{imagpi}Imaginary part of the subtracted $\phi$ meson selfenergy
in nuclear matter for several densities.}
\end{figure}

In Fig. \ref{imagpi} we show the imaginary part of the subtracted $\phi$
selfenergy for different densities from $\rho_0 /4$ to $\rho_0$. The resulting
width of the $\phi$ meson grows as a function of the density, and after adding
the free width it reaches the value of around 30 MeV at the vacuum $\phi$ mass
for normal nuclear density. Most of the $\phi$ decay channels contributing to
the in-medium width have a smooth behaviour at the energies under consideration,
except for the $\phi \to K \Sigma^* h$ channel which, neglecting the $\Sigma^*$
width, has a threshold at 940 MeV. This threshold moves to higher energies as
density grows because of the repulsive potential felt by the kaons. The position
of this threshold could be used to constrain the kaon and $\Sigma^*$ optical
potentials.
The results are quite similar in size and shape to those of ref.
\cite{Oset:2001eg}, except for the more visible bump related to the $\Sigma^*$
channels. The shape of the imaginary part of the subtracted selfenergy also
agrees quite well with the results of \cite{Klingl:1998tm} although a larger
width of around 45 MeV is obtained in that work. The possible reasons of this
difference are related to the different approaches in the calculation of the
kaon selfenergies \cite{Oset:2001eg} and the consideration of the recoil factors
here.

\begin{figure}
\includegraphics[width=0.7\textwidth]{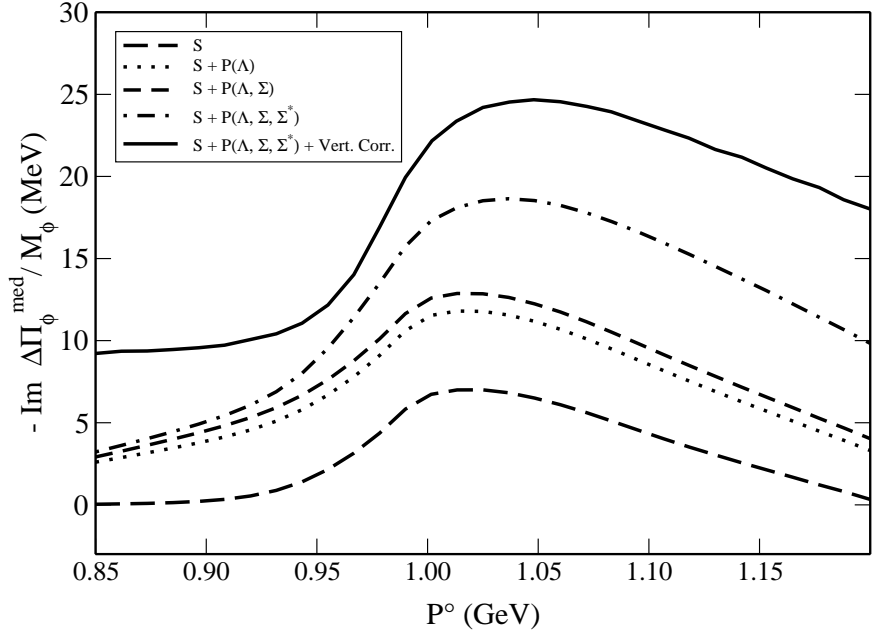}
\caption{\label{imagpieces}Contributions to the imaginary part of the subtracted
$\phi$ meson selfenergy in nuclear matter at $\rho=\rho_0$.}
\end{figure}

In Fig. \ref{imagpieces}, we show the $\phi$ subtracted width at normal nuclear
density obtained by successively considering different choices of antikaon
selfenergy and the inclusion of vertex corrections. In all cases we use the full
$S-$wave repulsive kaon selfenergy in the calculation. The lowest curve
corresponds to the $\phi$ subtracted medium width when the $\bar{K}$ propagator
incorporates only the $S-$wave $\bar{K}$ selfenergy. This selfenergy is mildly
attractive, compensating in part the repulsion felt by the kaon. Its imaginary
part comes from the kaon decaying into $\pi \Lambda h$ and $\pi \Sigma h$ and
therefore this corresponds to the $\phi$ decay channels $\phi \to K \pi \Lambda
h$ and $\phi \to K \pi \Sigma h$. 

In the next curves, we consider additionally the $P-$wave $\bar{K}$ selfenergy
due to the kaon coupling to a $\Lambda h$, $\Lambda h\,+\,\Sigma h$
and $\Lambda h\,+\,\Sigma h\,+\,\Sigma^* h$ excitations. We find a quite small
contribution from the $\Sigma$ and sizeable contributions from the 
$\phi N \to K \Lambda$ and $\phi N  \to K \Sigma^*$ processes. As commented
above, the $\Sigma^*$ channel has its threshold around 940 MeV and therefore
only contributes strongly above this energy. The small contribution below the
threshold, visible in the figure, is due to the approximation of using a
constant $\Sigma^*$ width in the calculation. The $\phi N \to K \Lambda$ channel
has a much lower threshold and thus its contribution depends little on energy.
Finally, the solid line also incorporates the vertex corrections leading to a
further enhancement of the total width.
\begin{figure}
\includegraphics[width=0.7\textwidth]{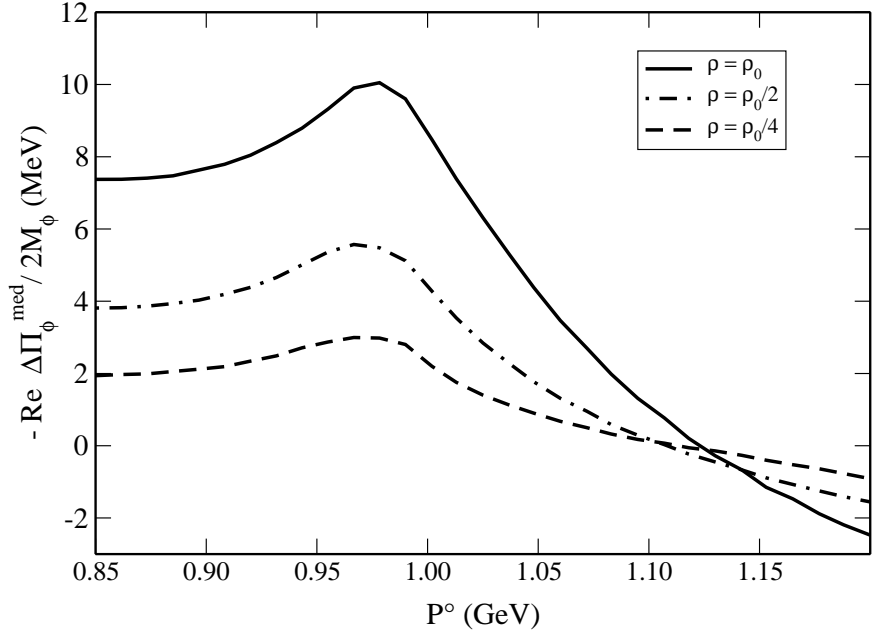}
\caption{\label{realpi}Real part of the subtracted $\phi$ meson selfenergy
in nuclear matter for several densities.}
\end{figure}

In Fig. \ref{realpi} we show the real part of $\Delta \Pi_{\phi}^{med}$ which
we find mildly attractive up to energies around 1.1 GeV. The change in the
$\phi$ mass at $\sqrt{s} = M_{\phi}$ is approximately 8 MeV to lower energies
at $\rho=\rho_0$. This small correction is in agreement with previous works
\cite{Klingl:1998tm,Kuwabara:1995ms}, although disagrees with naive scaling
expectations. There are recent experimental results at $T=0$ \cite{Ozawa:rk}
which do not observe any signature of in-medium modification of the $\phi$ mass
in the reaction $p A \to K^+ K^- X$. However, one should be careful in the
interpretation of the results without a detailed theoretical calculation. On
the one hand, the distortion of the final kaons and the initial protons could
force the reaction to occur at the nuclear surface and therefore at low
densities. On the other hand, due to the long $\phi$ lifetime and the large
average $\phi$ momentum a good part of the events could  correspond to $\phi$
decays in vacuum. Furthermore, for those events with a fast $\phi$ meson, the
use of a selfenergy obtained for a $\phi$ at rest is not justified.

In summary, we have studied the $\phi$ selfenergy for a $\phi$ at rest in a
cold symmetric nuclear medium, by considering the medium effects over the
tadpole and $K\bar K$ decay diagrams which dominate the $\phi$ vacuum
selfenergy. This has been done by dressing the kaons with $S-$ and $P-$wave
selfenergies. The $S-$wave antikaon selfenergy was obtained in a selfconsistent
coupled channel unitary calculation based on effective chiral Lagrangians in
which Pauli blocking, pion selfenergy and mean field potentials of the baryons
are taken into account. The $P-$ wave is driven by the coupling to hyperon-hole
excitations. In addition to these selfenergy insertions we have calculated a
set of vertex corrections required by gauge invariance. We obtain a small mass
shift of around 8 MeV and a width of around 30 MeV at normal nuclear density.
Thus, the main medium effect on  the $\phi$ meson is the growth of its width by
almost one order of magnitude  at normal nuclear density. Nevertheless, the
$\phi$ stays narrow enough to be  visible as an isolated resonance and then
give a clear signal in possible  experiments measuring dilepton or $\bar{K} K$
spectra.

\appendix
\section{\label{Ap1}On-shell matching of $\bar{K}NY$ static form factor}
\begin{figure}
\includegraphics[width=0.5\textwidth]{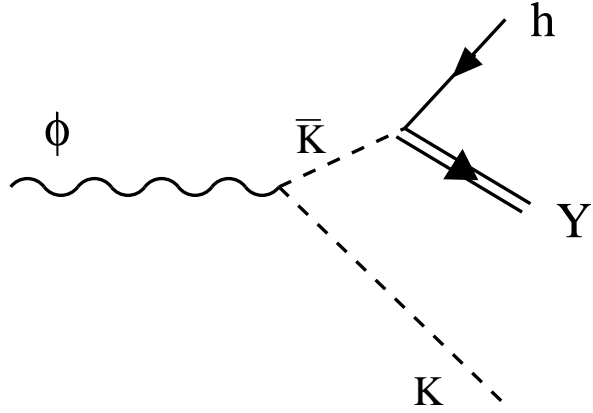}
\caption{\label{Yh_unitary_cut}$\phi$ meson decay into a $K$ meson and a $Y-h$
excitation.}
\end{figure}

In this work we have used the model of ref. \cite{Oset:2001eg} for the
$\bar{K}$ selfenergy. However, as we are interested in the real part of the
$\phi$ selfenergy and therefore we need to evaluate loop contributions, we have
introduced new $\bar{K}NY$ form factors and more precise recoil corrections. In
order to keep consistency with the coupling of previous calculations
\cite{Klingl:1998tm,Oset:2001eg} we match our couplings at the appropriate
kinematic conditions.  Let us consider the process depicted in Fig.
\ref{Yh_unitary_cut} in which a $\phi$ meson decays into a kaon and $Y-h$
excitation.
It corresponds to the unitary cut of the selfenergy diagram in Fig.
\ref{vc_graphs}a and relates to the $\phi$-nucleon scattering process $\phi N \to KY$.
In the situation of a $\phi$ at rest respect to the nuclear medium the energies
and momenta carried by the on-shell $K$ and the off-shell $\bar{K}$ are given by
\ba
\label{energybalance}
E_K = \frac{(M_N + M_{\phi})^2 + m_K^2 - M_Y^2}{2 (M_N + M_{\phi})},
\nonumber \\
E_{\bar{K}} = M_{\phi} - E_K,
\nonumber \\
|\vec{q}|_K = |\vec{q}|_{\bar{K}} = \sqrt{E_K^2 - m_K^2}.
\ea
The matching proceeds by demanding the following equality to be satisfied,
\be
\label{match}
C^2 \bigg ( \frac{\Lambda ^2}{\Lambda ^2 + |\vec{q}|_K^2} \bigg )^4
f_Y^2(E_{\bar{K}},|\vec{q}|_{\bar{K}})
= \bigg ( \frac{\Lambda^2}{\Lambda^2 - E_{\bar{K}}^2 +
|\vec{q}|_K^2} \bigg )^4 f_{\textrm{\tiny{0}} \, \, Y}^2(E_{\bar{K}}) ,
\ee
what fixes the value of the scaling constant $C$ in terms of the chosen cut-off
parameter $\Lambda = 1.05$ GeV \cite{Klingl:1998tm,Oset:2001eg}. We set $M_Y$
to an average value of 1200 MeV, $f_Y^2$ is given in eq. (\ref{RF_B}) and
$f_{\textrm{\tiny{0}} \, \, Y}^2 = (1-E_{\bar{K}}/(2M_Y))^2$ is the recoil
factor of ref. \cite{Oset:2001eg}. We get a scaling constant value of $C =
1.04$.

\section{\label{Ap2}$S-$wave gauging terms}
\begin{figure}
\includegraphics[width=0.6\textwidth]{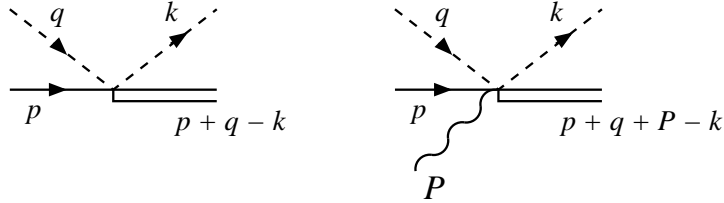}
\caption{\label{vertex_gauge_S}$S-$wave kaon-nucleon vertex and the corresponding $\phi$
contact coupling.}
\end{figure}

We evaluate here the selfenergy contributions from diagrams in Fig.
\ref{vc_graphsII}(b-d). The necessary Feynman rules for the vertices involved
(Fig. \ref{vertex_gauge_S}) are given by
\ba
\label{MMNY}
-i t_{\bar{K}NMY} = i \frac{1}{4f^2} C_{\textrm{\tiny i,f}} \gamma^{\mu}
(q+k)_{\mu}
\nonumber \\
-i t_{\bar{K}NMY\phi} = i g_{\phi} \frac{1}{4f^2} C_{\textrm{\tiny i,f}}
\gamma^{\mu} \epsilon_{\mu} (\phi) ,
\ea
where the $C_{\textrm{\tiny i,f}}$ coefficients, related to the initial and
final states, can be found in \cite{Oset:1998it}. We use a non-relativistic
reduction of these vertices which is enough to determine the size of
the imaginary part. We only keep the temporal components in the slash appearing
in $t_{\bar{K}NMY}$ and the ${\cal O}(1/M_B)$ terms in $t_{\bar{K}NMY\phi}$. In
addition we neglect terms linear in the hole momentum. We select the case of an
intermediate $\pi^0 \Lambda$ state, which is the one less disfavored in phase
space. Under these assumptions we find
\ba
\label{VC1_Swave}
\Pi_{\phi}^{VC1'} (P^0,\rho) = - i g_{\phi}^2 \frac{1}{128 f^4 M_{\Lambda}
\pi^2} \frac{1}{P^0} \bigg \lbrace \mathcal{P} \int_0^{\infty} dq
\frac{\vec{q}\,^2}{\omega_K(q)} \bigg \lbrack
\frac{I(-P^0-\omega_K(q),|\vec{q}|)}{P^0+2\omega_K(q)} +
\nonumber \\
\frac{I(-\omega_K(q),|\vec{q}|)}{P^0-2\omega_K(q)} \bigg \rbrack - i \pi
 \frac{|\vec{q}_{on}|}{2} \, I(-P^0/2,|\vec{q}_{on}|) \, \theta (P^0-2
m_k) \bigg \rbrace ,
\ea
with
$I(q^0,|\vec{q}|) = \frac{-i}{(2\pi)^2} \int_0^{\infty} dk
\, \vec{k}\,^2 \, \frac{q^0+P^0-\omega_{\pi}(k)}{2 \omega_{\pi}(k)}
\int_{-1}^{+1} du \, (\vec{q}\,^2-\vec{q} \cdot \vec{k}) \,
U_{\Lambda}(q^0+P^0+\omega_{\pi}(k),\vec{q}-\vec{k};\rho /2)$, 
$\sigma(P^0) = \sqrt{1-4 m_K^2 / (P^0)^2}$, 
$|\vec{q}_{on}| = \sigma(P^0) \, P^0 /2$, 
$\omega_K(q) = \sqrt{\vec{q}\,^2+m_K^2}$, $\omega_{\pi}(k) =
\sqrt{\vec{k}\,^2+m_{\pi}^2}$, and
$u \equiv \frac{\vec{q} \cdot \vec{k}}{|\vec{q}| \cdot |\vec{k}|}$
for diagrams (b) and (c), and
\be
\label{VC2_Swave}
\Pi_{\phi}^{VC2'} (P^0,\rho) = i g_{\phi}^2 \frac{3}{256 f^4 M_{\Lambda}^2}
\frac{1}{2\pi^2} \int_0^{\infty} dq \frac{\vec{q}\,^2}{2\omega_K(q)}
\widetilde{I} (-\omega_K(q),|\vec{q}|) ,
\ee
with
\begin{displaymath}
\label{VC2_Swave_detail}
\widetilde{I}(q^0,|\vec{q}|) = \frac{-i}{(2\pi)^2} \int_0^{\infty} dk \,
\frac{\vec{k}\,^2}{2\omega_{\pi}(k)} \int_{-1}^{+1} du \, (\vec{q}-\vec{k})^2
\, U_{\Lambda}(q^0+P^0+\omega_{\pi}(k),\vec{q}-\vec{k};\rho /2)
\nonumber
\end{displaymath}
for diagram (d). The imaginary parts of $\Pi_{\phi}^{VC1'}$ and
$\Pi_{\phi}^{VC2'}$ are finite and very small compared to the contributions of
meson selfenergy insertions and $P-$wave vertex corrections. The
real parts are divergent and one has to regularize both the $\pi - Yh$ loop and
the external kaon loop. Under reasonable momentum cut-offs ($1 - 2$ GeV) we
find that the real parts are of the same order as the imaginary parts and thus
negligible compared to the leading contributions.

\begin{acknowledgments}
We thank E. Oset for fruitful discussions. 
We acknowledge partial financial support from the DGICYT under
contract BFM2000-1326. D.~C. thanks financial support from MCYT.
\end{acknowledgments}

\end{document}